# Tuning a binary ferromagnet into a multi-state synapse with spin-orbit torque induced plasticity

Yi Cao[1], Andrew Rushforth[2], Yu Sheng[1], Houzhi Zheng[1,3], Kaiyou Wang[1,3,*]

[1]*State Key Laboratory for Superlattices and Microstructures, Institute of Semiconductors, Chinese Academy of Sciences, Beijing 100083, P. R. China*

[2]*School of Physics and Astronomy, University of Nottingham, Nottingham NG7 2RD, United Kingdom*

[3]*College of Materials Science and Opto-Electronic Technology, University of Chinese Academy of Science, Beijing 100049, P. R. China*

\* Corresponding emails: kywang@semi.ac.cn

**Abstract**

**Inspired by ion-dominated synaptic plasticity in the human brain, electronic artificial synapses for neuromorphic computing adopt charge-related quantities as their weights. Despite the existing charge derived synaptic emulating strategies, schemes of controlling electron spins in binary ferromagnetic devices have also attracted considerable interest due to their advantages of low energy consumption, unlimited endurance, and favorable complementary metal–oxide–semiconductor (CMOS) compatibility. However, a generally applicable method of tuning the binary ferromagnet into a multi-state memory with pure spin-dominated synaptic plasticity in the absence of an external magnetic field is still missing. Here, we show how synaptic plasticity of a perpendicular ferromagnetic layer (FM1) can be**

**obtained when it is interlayer-exchange-coupled by another in-plane ferromagnetic layer (FM2), where a magnetic-field-free current-driven multi-state magnetization switching of FM1 in the Pt/FM1/Ta/FM2 structure is induced by spin-orbit torque. We use current pulses to set the perpendicular magnetization state (represented by the anomalous Hall resistance) which acts as the synapse weight, and demonstrate spintronic implementation of the excitatory/inhibitory postsynaptic potentials and spike timing-dependent plasticity. This functionality is made possible by the action of the in-plane interlayer exchange coupling field which leads to broadened, multi-state magnetic reversal characteristics. Numerical simulations, combined with investigations of a reference sample with a single perpendicular magnetized Pt/FM1/Ta structure, reveal that the broadening is due to the in-plane field component tuning the efficiency of the spin-orbit-torque to drive domain walls across a landscape of varying pinning potentials. The conventionally binary FM1 inside our Pt/FM1/Ta/FM2 structure with inherent in-plane coupling field is therefore tuned into a multi-state perpendicular ferromagnet and represents a synaptic emulator for neuromorphic computing, paving a substantial pathway towards the combination of spintronics and synaptic electronics.**

The source of human memory and learning lies in the plasticity of brain synapses, where the degree of connections between neurons is set by the synaptic strength (or weight), which is much more efficient than conventional computational systems in solving complex problems.[1] Inspired by this, the field of neuromorphic computing, that emulates synaptic learning functions by a single electronic device and then integrates the synaptic devices to neuron circuits, has attracted significant interest, resulting in inspirations in operating principles, algorithms and architectures.[2-4] The demonstration of synapse-like electronic devices are mainly focused on nonvolatile charge memory technologies, including field effect transistors, phase change memory, resistance change memory, ferroelectric switches, etc. These involve manipulating ions and/or electrons/holes and setting the resulting multi-state conductivity as the weight.[5-11]

In addition to the charge-based devices, devices based on the spin degree of freedom, such as magnetoelectric coupled memtransors, spin-transfer torque (STT) memristors, and spin-orbit torque magnetic bilayers have also been suggested for implementation in synaptic electronics recently.[12-20] Stochastic synapses that utilize the intrinsic time randomness of STT reversal in a single magnetic tunnel junction (MTJ) have already shown their potential for spiking neural network (SNN) applications.[21,22] For the implementation of conventional multi-state synaptic plasticity, current-driven magnetic domain wall motion has been employed by the spintronic community[16-19]. However, since the exhibited magnetic states of a ferromagnet are often binary, or consist of multiple states with no control over stochastic switching fields,[17] a generally applicable method that allows a binary ferromagnet to exhibit tunable multi-state current-driven switching behavior is strongly desired.

Unlike current-driven devices employing the STT, it is not necessary for the electric current to pass directly through the ferromagnetic layers in devices employing spin-orbit torques (SOTs). Therefore, current-induced magnetization switching by SOTs can have the advantages of simpler device design, and more reliable and energy efficient operation.[23] In heavy metal (HM)/ferromagnetic metal (FM) bilayer systems, the origins of the SOTs are the spin Hall effect (SHE) in the HM and/or the Rashba effect from interfacial inversion asymmetry.[24-26] When the magnetization reversal mechanism involves the nucleation and driving of domain walls by the SOTs, the presence of the Dzyaloshinskii-Moriya interaction (DMI) arising at the heavy metal/ferromagnet interface necessitates the application of a static in-plane magnetic field to realize current induced deterministic switching of perpendicular magnetization.[27] Without an external magnetic field, switching assisted by equivalent in-plane magnetic fields has been demonstrated using a wedge oxide capping layer, a polarized ferroelectric substrate, and stacks with exchange bias or interlayer exchange coupling using antiferromagnetic layers.[20,28-32] Considering that antiferromagnets have strong exchange interaction between neighboring magnetic atoms and thus small domains, spin-orbit torque induced gradual switching behaviors are observed not only in an antiferromagnet itself, but also in the neighboring ferromagnetic layer interfaced with the antiferromagnet, respectively.[20,33] Inspired by this, Olejník et al[34] and Borders et al[35] respectively developed an antiferromagnetic CuMnAs multi-level memory cell, and a network consisting of 36 antiferromagnet/ferromagnet bilayer devices with magnetic-field-free multi-state SOT switching behaviors for associative memory operations, recently. However, for a ferromagnetic system it is still challenging to experimentally control the number of magnetic switching states (from binary to multiple states) in order to enable it as a spintronic synapse.

Here, we demonstrate magnetic-field-free SOT-induced multi-state synaptic plasticity in a Pt/FM1/Ta/FM2 structure, where the layer with perpendicular magnetic anisotropy (PMA), FM1, is antiferromagnetically coupled with the layer with in-plane magnetic anisotropy (IMA), FM2. Using imaging Kerr microscopy measurements and numerical simulations, we show that the magnetization reversal proceeds via domain wall motion, with the exchange field from the top magnetic layer tuning the efficiency of the SOT-induced domain wall motion as a function of the strength of the pinning potential. This method of simply coupling the FM1 with the FM2 results in a range of accessible magnetic states in the FM1 and enables emulation of the synaptic functionality. Based on these spintronic devices, we demonstrate synaptic functionalities including excitatory postsynaptic potential (EPSP), inhibitory postsynaptic potential (IPSP), and spike timing-dependent plasticity (STDP), which could be important for developing neuromorphic computing.[36-38]

**Device preparation and measurement configuration**

A Pt/FM1/Ta/FM2 system with stacks of Ta(0.5)/Pt(3)/Co(1.3)/Ta(1.2)/Co(4)/AlO$_x$(2) and a reference Pt/FM1/Ta system with stacks of Ta(0.5)/Pt(3)/Co(1.3)/Ta(1.2)/AlO$_x$(2) (from the substrate side, thickness in nm) were deposited on Si/SiO$_2$ substrates by magnetron sputtering. After deposition, the Pt/FM1/Ta/FM2 films were processed into Ta(0.5)/Pt(3) hall bars with a 6 μm-wide Co(1.3)/Ta(1.2)/Co(4)/AlO$_x$(2) square pillar on top of the crossing area of each Hall bar. For the Pt/FM1/Ta/FM2 system, magneto-optical Kerr characterization of the pillar shows rectangular hysteresis loops in both out-of-plane and in-plane directions (see Supplementary S1), indicating that the thinner Co(1.3 nm) layer exhibits PMA and the thicker Co(4 nm) layer exhibits IMA, which respectively serve as the free layer and the fixed layer. The Pt/FM1/Ta reference system was processed into the same device configuration, i.e. Ta(0.5)/Pt(3) Hall bars, but with a 6 μm-wide (and also a 20 μm-wide) Co(1.3)/Ta(1.2)/AlO$_x$(2) square pillar at the cross of each Hall bar.

Measurements of both the magneto-optic Kerr effect and the Hall resistance ($R_H$) from the anomalous Hall effect (AHE) were used to evaluate the perpendicular magnetization component ($m_z$) of the Co(1.3 nm) free layer in the pillars. As shown in Figure 1a, the input currents for the AHE measurement and for the SOT-induced switching were applied along the $x$-direction, while the AHE voltage ($V_H$) was detected at the other channel with +$y$-direction. $R_H$ was obtained by dividing $V_H$ with a low measurement d.c. current of 100 μA. Note, that with this measurement geometry, the sign of $R_H$ is in accordance with the direction of $m_z$, i.e. positive (negative) $R_H$ represents +$z$ (-$z$) directed moment.

## Demonstration of spintronic EPSP/IPSP and STDP

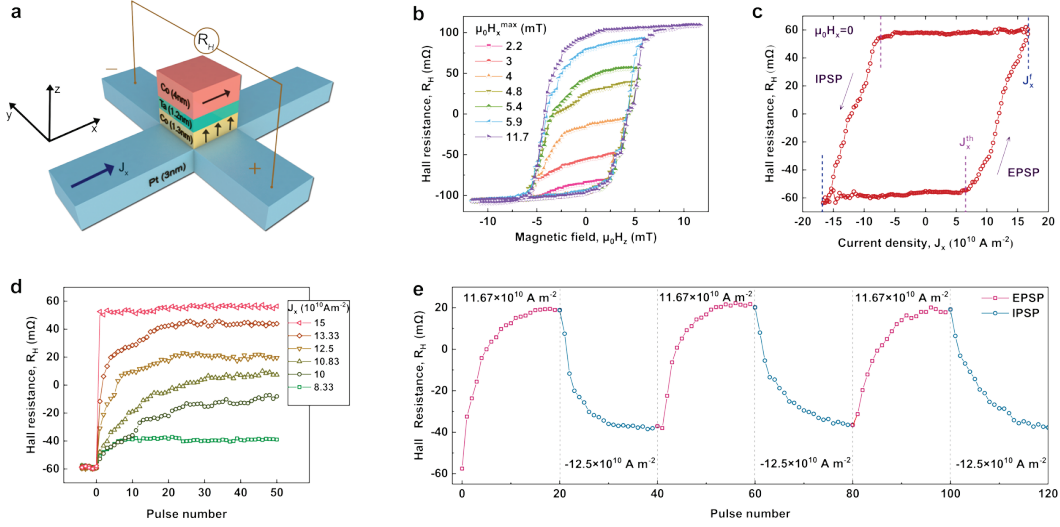

**Figure 1 Device structure, magnetic switching properties, and excitatory/inhibitory synaptic behaviors. a,** Schematic of the Pt/FM1/Ta/FM2 double magnetic layered system and the AHE/SOT measurement configuration with definition of x-y-z coordinates. The device consists of a Ta(0.5 nm)/Pt(3 nm) Hall bar with a 6 μm-wide Co(1.3 nm)/Ta(1.2 nm)/Co(4 nm)/AlO$_x$(2 nm) square pillar at the cross, where the FM1 Co(1.3 nm) layer and the FM2 Co(4 nm) layer exhibit PMA and IMA, respectively. The magnetization of the FM2 Co(4 nm) layer is pre-magnetized along the +x-direction. **b,** $R_H$-$H_z$ loops with $\mu_0H_z$ scanning from -11.7 mT to $\mu_0H_z^{max}$ and then back to -11.7 mT. **c,** The $R_H$-$J_x$ loop with no external field. A sequence of pulses with scanning magnitude from -16.67×10$^{10}$ A m$^{-2}$ to 16.67×10$^{10}$ A m$^{-2}$ and then back to -16.67×10$^{10}$ A m$^{-2}$ is applied. The $J_x^{th}$ and the $J_x^f$ denote the threshold and the finishing current densities for the SOT-induced switching, respectively. **d,** $R_H$ as a function of the pulse number for sequences of pulses with constant magnitude $J_x$. **e,** Evolution of the EPSP/IPSP (i.e. $R_H$) by applying a sequence of pulse strings with magnitude of 11.67×10$^{10}$ A m$^{-2}$ and -12.5×10$^{10}$ A m$^{-2}$. Before apply each sequence of pulses for **d** and **e**, an initialization pulse with a magnitude of -16.67×10$^{10}$ A m$^{-2}$ is applied to the channel. The duration of every pulse is 10 ms, and every $R_H$ is measured 2 seconds after the application of each pulse. For the AHE detection of $R_H$ in **b-e**, a small +x directed d.c. current with fixed magnitude of 100 μA is used as the current source.

We characterize the Pt/FM1/Ta/FM2 device by examining its $R_H$-$\mu_0H_z$ loops as a function of maximum applied magnetic field $\mu_0H_z^{max}$. Prior to the measurements, a 20 mT external magnetic field was used to pre-magnetize the magnetization of the in-plane Co(4 nm) layer along the +x-direction. Then, the in-plane magnetic field was removed. Figure 1b shows the $R_H$-$\mu_0H_z$ loops measured with the perpendicular field scanning from -11.7 mT to several different $\mu_0H_z^{max}$, and then back to -11.7 mT, where the $\mu_0H_z^{max}$ varies from smaller to larger than the saturation field ($\mu_0H_z^s$, around 7 mT). Instead of showing binary states, the loops enlarge gradually with the increasing $\mu_0H_z^{max}$ before it reaches the saturation field, indicating controllable magnetization reversal processes in our structure. Next, we investigate the response of $R_H$ to $J_x$ under zero magnetic field, where the in-plane Co(4 nm) fixed layer is expected to provide an assisting field for SOT-induced switching of the perpendicular Co(1.3 nm) free layer.[39] After setting the device to a negative maximum $R_H$ state, we applied a sequence of 10 ms-long pulses with scanning magnitude from -16.67×10$^{10}$ A m$^{-2}$ to 16.67×10$^{10}$ A m$^{-2}$ and then back to -16.67×10$^{10}$ A m$^{-2}$. A low d.c. current of 100 μA was used to monitor $R_H$ after each pulse, and the measured $R_H$-$J_x$ loop is plotted in Figure 2c. A change in $R_H$ with $J_x$ is observed as expected, indicating that the perpendicular magnetization in the Co(1.3 nm) free layer is reversed solely by the current-induced torque. The threshold switching current density ($J_x^{th}$) and the finishing switching current density ($J_x^f$) are defined as the absolute values of $J_x$ at which the change of $R_H$ begins and finishes, which is around 7×10$^{10}$ A m$^{-2}$ and 16.67×10$^{10}$ A m$^{-2}$ for current applied along +x-direction in Figure 2c. Similar to the $R_H$-$\mu_0H_z$ loops, the $R_H$-$J_x$ loop also exhibits a gradual reversal behavior, showing that the multiple nonvolatile perpendicular magnetic states in our device can be obtained through current-induced switching. The sense of rotation of the $R_H$-$J_x$ loop is reversed if FM2 is pre-magnetized along the -x-direction (see Supplementary S2). This flexible feature of adjustable current-induced switching directions in our multi-state magnetic system will bring additional functionalities when combined with the synaptic implementations discussed below. The anticlockwise and clockwise sense of field free current-induced magnetization switching with pre-magnetization of the top FM2 layer along the +x and -x directions respectively, suggests that there is in-plane antiferromagnetic exchange coupling between FM1 and FM2.[39,40] The magnitude of this coupling field ($\mu_0H_x^c$) is at an optimum value of around -180 mT for our multilayer stacks, as illustrated by the $R_H$-$J_x$ characteristics under various magnetic fields along the +x-direction ($\mu_0H_x$) (see Supplementary S3).

This multi-magnetic-state switching behavior under current pulses can be analogous to the information transmission characteristics of biological synapses. A synapse is a conjunction of two neuron cells, named pre-neuron and post-neuron. Under an external stimulus, spikes or action potentials from the pre-neuron are transmitted through the synapse to the post-neuron and generate excitatory postsynaptic potentials (EPSP) or inhibitory postsynaptic potentials (IPSP), together with synaptic weight updates.[36] Here, the artificial synapse characteristics based on our Pt/FM1/Ta/FM2 structure can be achieved by regarding the perpendicular magnetization component, $m_z$ as the synaptic weight. The applied current pulses play the roles of the pre- and post-spikes that act on our artificial synapse to modulate $m_z$, which is assessed by the detection of $R_H$. Therefore, the increasing and decreasing of $R_H$, as shown in Figure 1c, indicates the synaptic plasticity of EPSP and IPSP, respectively. In addition to the pulse magnitude-induced EPSP/IPSP, we also examined the effect of pulse numbers on the switching of $R_H$ at different fixed pulse magnitudes. As shown in Figure 1d, sequences of 50 pulses with constant magnitude ($J_x$) and duration of 10 ms are applied to the channel, and $R_H$ is measured 2 seconds after each pulse. Before the application of each sequence, an initialization pulse with magnitude of -16.67×10$^{10}$ A m$^{-2}$ was applied to align the magnetization in the -z-direction. With increasing pulse number, $R_H$ initially increases quickly and then gradually reaches a relatively stable value for $J_x$ ranging from 8.33×10$^{10}$ A m$^{-2}$ to 13.33×10$^{10}$ A m$^{-2}$, where the stable value increases as a function of the pulse magnitude. For $J_x$ = 15×10$^{10}$ A m$^{-2}$, $R_H$ jumps directly from the minimum to approximately the maximum value by the first pulse, which is because the applied pulse is close to $J_x^f$. By applying a sequence of 6 alternating pulse strings (20 pulses per string) with moderate $J_x$ of 11.67×10$^{10}$ A m$^{-2}$ and -12.5×10$^{10}$ A m$^{-2}$, consecutive EPSPs and IPSPs are induced only by pulse numbers, as shown by the patterns of multiple increasing and decreasing $R_H$ values in Figure 1e. Since sequences with identical pulses required simpler circuit design than those with varying magnitude, the EPSP/IPSPs used in Figure 1e will be more practical to implement in neural network systems than the EPSP/IPSPs shown in Figure 1c.[12]

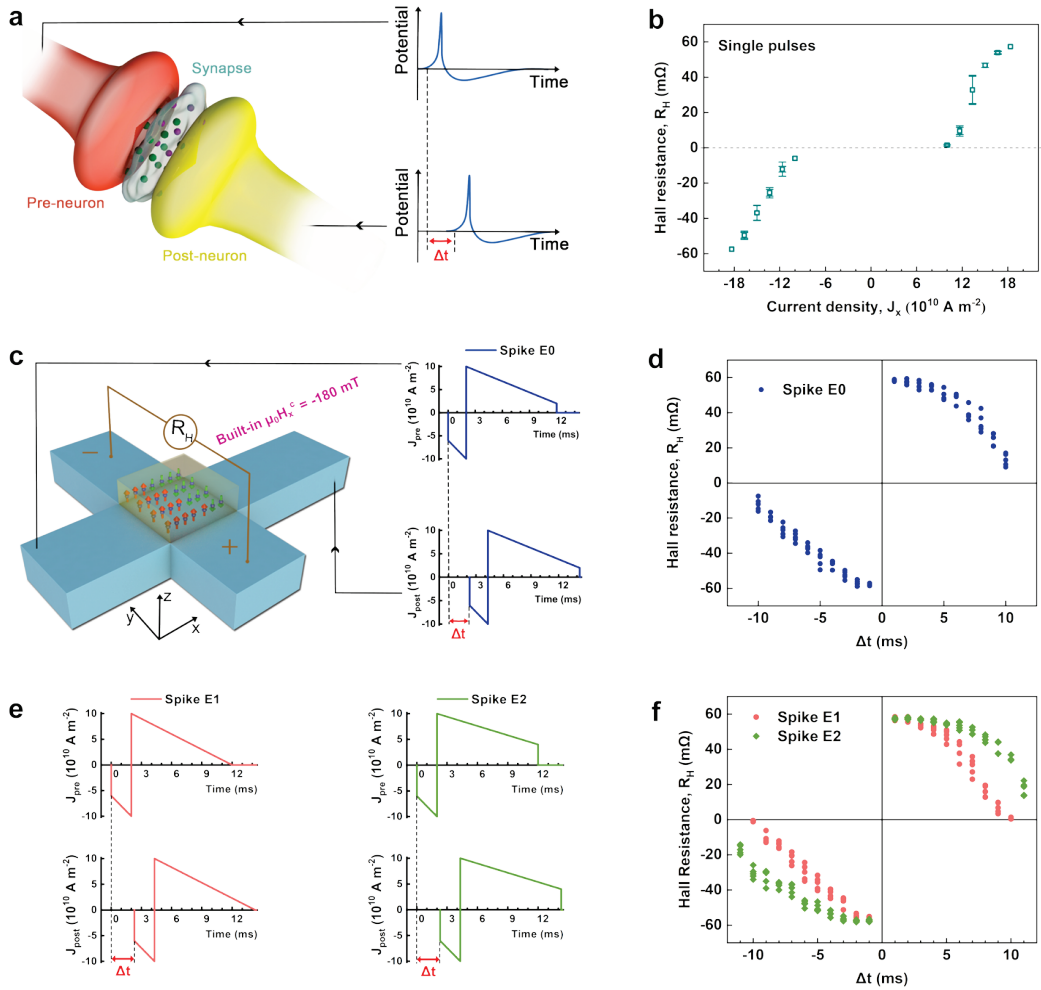

**Figure 2 Field-free SOT-induced STDP in Pt/FM1/Ta/FM2 system. a,** Sketch of a biological synapse sandwiched between pre- and post-neurons, where the synaptic transmission is modulated by the causality (Δt) of neuron stimuli which controls the ion channels. **b,** $R_H$ as a function of $J_x$ for the Pt/FM1/Ta/FM2 device (initialized in the $R_H \approx 0$ state) when only one current pulse with duration of 1 ms is applied. **c,** Sketch of our Pt/FM1/Ta/FM2 artificial synapse. The $m_z$ (i.e. $R_H$) is modulated by the causality (Δt) of the engineered spikes (named Spike E0) from both ends of the x channel. **d,** STDP window (i.e. the $R_H$-Δt distribution) with the spikes shown in **c** (blue spike curves). **e,** another two spikes with different decay slopes named Spike E1 (pink spike curves) and Spike E2 (green spike curves). **f,** STDP window generated with spikes E1 and E2. Before the application of every single pulse or mixed spike in **b**, **d**, and **f**, the sample is switched to an almost zero $m_z$ state with $R_H \approx 0$. The AHE

measurement with a small +x directed d.c. current source of 100 μA is taken more than 2 seconds after the singe pulse or mixed spike ends. This procedure (set $R_H \approx 0$ → apply pulse/spike → measure $R_H$ → set $R_H \approx 0$) repeats at least five times for each $J_x$ or $\Delta t$. The spread of $R_H$ data points is represented in the forms of error bars (in **b**) and scatter diagrams (in **d** and **f**), respectively.

Figure 2a illustrates STDP. A long-term depression (LTD) or long-term potentiation (LTP) of the synaptic weight occurs when the post-neuron spikes just before ($\Delta t < 0$) or after ($\Delta t > 0$) the pre-neuron spikes, respectively, where the typical STDP window of a biological synapse can be found in Figure 7 of ref 37.[37] It is also characterized by the $|\Delta t|$ dependence of the synaptic potentiation/depression magnitudes, i.e. closer spikes bring stronger stimulation. A biological synapse realizes such plastic functions by regulating the ion concentrations inside it,[41] whereas a spintronic synapse may adopt the changes in magnetic state during the current controlled switching process as an analogue. The existence of separate $J_x^{th}$ and $J_x^f$ makes it possible to implement STDP in our device. Changes in $R_H$ induced by single current pulses, as a function of the pulse magnitude $J_x$ are shown in Figure 2b. Before the application of each single pulse, the sample was reset to an almost zero $m_z$ state ($R_H \approx 0$). For $J_x > J_x^{th}$ or $-J_x > -J_x^{th}$, an increase or a decrease of $R_H$ can be induced by the single pulse, and higher $|J_x|$ brings greater change in $R_H$ until $|J_x|$ approaches $|J_x^f|$ (note that $|J_x^f|$ for single pulse here is slightly larger than that for consecutive pulses in Figure 1c). Based on this $J_x$ dependence of $R_H$ under single current pulse, spikes as shown in Figure 2c are designed for the demonstration of STDP in our device. The pre-spike and post-spike share the same waveform but are opposite in polarity. In this way, although the current density of a single pre- or post-spike never exceeds $J_x^{th}$, the overlapped waveform will produce a short over threshold peak, whose sign and magnitude depend on the $\Delta t$. Results of the pre-and post-spikes acting on the channel of our device with $R_H \approx 0$ are shown in Figure 2d, where a field-free SOT-induced STDP window is obtained. The distribution of the weakened (LTD, when $\Delta t < 0$) and the strengthened (LTP, when $\Delta t > 0$) Hall resistances show decent fidelity with two nearly linear $R_H$-$\Delta t$ responses in the first and the third quadrants, as well as tolerable $R_H$ deviations for repeated measurements. The form of a STDP window is important because different STDP forms can emulate different types of neuron activities that vary significantly according to the location of the synapses in a brain, and the outcome of a certain learning procedure in a spike neural network is highly dependent on the STDP form of an artificial synapse.[6,11] In Figure 2e, we then modified the spike shapes with two adjusted decay slopes and obtained corresponding STDP windows with distinct forms, as shown in Figure 2f. This further illustrates that the demonstrated spintronic synapse could be practically used for spike-based neuromorphic computing applications with the capability of implementing versatile STDP forms, including the exponential STDP form,[42] via simply modulating the spike shapes (see the simulations in Supplementary S4 for detail).

# Correlation between multi-state $m_z$ and $\mu_0H_x$

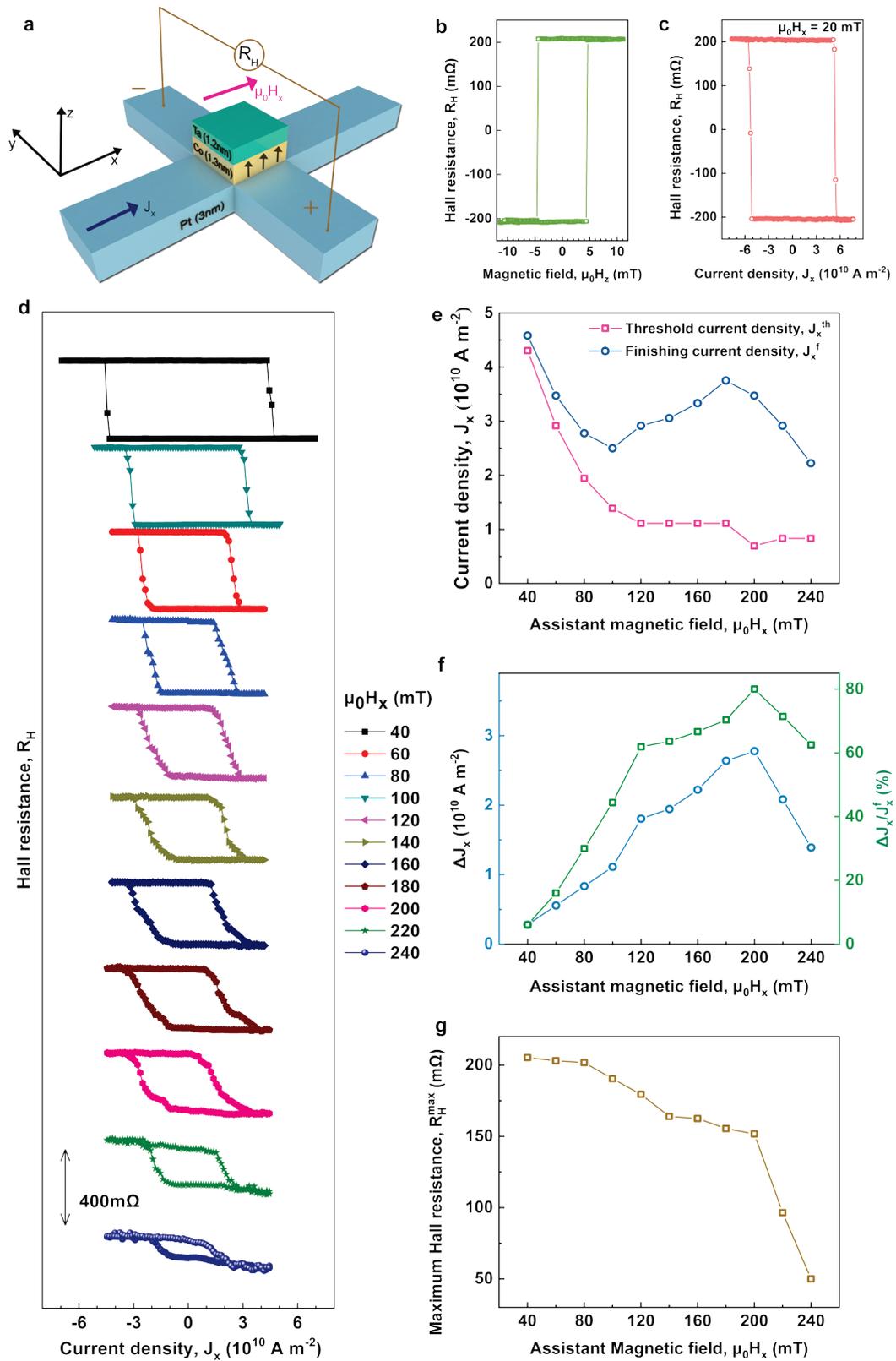

**Figure 3 SOT switching of Pt/FM1/Ta reference sample. a,** schematic of the Pt/FM1/Ta reference device that consists of a Ta(0.5 nm)/Pt(3 nm) Hall bar with a 6 μm-wide Co(1.3

nm)/Ta(1.2 nm)/AlO$_x$(2 nm) square pillar at the cross. During the measurements, an external assistant magnetic field $\mu_0H_x$ is applied along the +x-direction. **b,** $R_H$-$\mu_0H_z$ loop with $\mu_0H_x = 0$. **c,** Pulse current-driven $R_H$-$J_x$ SOT switching loop with $\mu_0H_x = 20$ mT. **d,** Pulse current-driven $R_H$-$J_x$ SOT switching loops with various $\mu_0H_x$ from 40 mT to 240 mT. Loops are shifted vertically for clarity. The applied pulse sequences for **c** and **d** are pulse strings with each duration of 10 ms and scanning magnitude between the maximum $\pm J_x$, and all $R_H$ data point are measured 2 seconds after the application of each pulse. For the AHE detection of $R_H$ in **b-d**, a small +x directed d.c. current with fixed magnitude of 100 μA is used as the current source. **e,** Evolution of $J_x^{th}$ and $J_x^f$ with the increasing $\mu_0H_x$. **f,** $\mu_0H_x$ dependence of the current controllable region $\Delta J_x$ ( = $J_x^{th}$ - $J_x^f$, the blue circles) and the current controllable proportion $\Delta J_x/J_x^f$ (the green spots). **g,** Maximum anomalous Hall resistance $R_H^{max}$ as a function of $\mu_0H_x$.

To understand the origin of the multi-state magnetization reversal of FM1 in the Pt/FM1/Ta/FM2 system, we will focus on a Pt/FM1/Ta reference system without the FM2 layer. The measured $R_H$-$H_z$ loop for the Hall bar with 6 μm-wide square pillar, shown in Figure 3b, exhibits perfect rectangularity, representing binary $m_z$ states in the Pt/FM1/Ta reference structure with good PMA. This is distinct from the multi-state magnetization reversal observed in the Pt/FM1/Ta/FM2 system. Without an in-plane external magnetic field applied, no deterministic current-induced magnetization switching is observed. With a small external in-plane magnetic field, $\mu_0H_x = 20$ mT, applied in the +x-direction, a very sharp current pulse induced clockwise $R_H$-$J_x$ loop is observed (Figure 3c). Interestingly, with increasing the external magnetic field from 40 mT to 240 mT, the $R_H$-$J_x$ loops, as shown in figure 3d become more parallelogram-like. The dependence of $J_x^{th}$ and $J_x^f$ on $\mu_0H_x$ are plotted in Figure 4e. $J_x^{th}$ decreases continually with increasing $\mu_0H_x$, while the variation of $J_x^f$ does not obey the $J_x^{th}$-$H_x$ trend well, instead showing a relatively slowly decreasing $J_x^{th}$-$H_x$ relationship as $\mu_0H_x$ increases from 40 mT to 100 mT, an increasing $J_x^{th}$-$H_x$ dependence when $\mu_0H_x$ ranges from 100 mT to 180 mT, and a decreasing $J_x^{th}$-$H_x$ relation again for $\mu_0H_x$ larger than 180 mT. As a consequence, an increasingly current-controllable region (defined as $\Delta J_x = J_x^{th} - J_x^f$) with the growing $\mu_0H_x$ is obtained, as shown in Figure 3f. To quantify the degree of the current control of the multiple magnetic states, we introduce $\Delta J_x/J_x^f$ to describe the fraction of current-controllable region during the applied current pulse rising from zero to $J_x^f$. As shown in Figure 3f, peaks in $\Delta J_x$ and $\Delta J_x/J_x^f$ correspond with $\mu_0H_x$ around 180 mT and 200 mT, suggesting our Ta(0.5)/Pt(3)/Co(1.3)/Ta(1.2)/Co(4)/AlO$_x$(2) system with built-in $\mu_0H_x^c \approx$ -180 mT is the optimized structure for synaptic emulation in terms of the availability of multiple magnetization states. In the Supplementary S5 we present numerical calculations which reveal that the increase of $\Delta J_x$ with $\mu_0H_x$ arises from the tuning of the efficiency of the SOT to drive domain wall motion in a pinning potential by $\mu_0H_x$. In addition, $\mu_0H_x$ tilts $m_z$ within the up/down domains, leading to a reduction in the maximum values of $R_H$ ($R_H^{max}$). Figure 3g shows $R_H^{max}$ as a function of $\mu_0H_x$. An external field with $\mu_0H_x >$ 200 mT causes overwhelming loss of PMA, and coincides with the drop of $\Delta J_x$ and $\Delta J_x/J_x^f$, resulting in a loss of accessible magnetic states.

In this way, we have successfully tuned an originally binary perpendicular ferromagnet FM1 into a multi-state one by simply coupling it with another in-plane ferromagnet FM2 via a normal metal spacer layer (in our case, Ta). The particular $m_z$ among the multiple perpendicular magnetic states of the FM1 inside our Pt/FM1/Ta/FM2 structure can be controlled by spin-orbit torque induced current-driven magnetic switching without assistance of any external magnetic field, making it capable of performing synaptic plasticity of EPSP, IPSP, and STDP. Note that the magnitude of the interlayer exchange coupling field can be manipulated by a fully electrical method such as ionic liquid gating, which would effectively change the functions of $R_H$ versus single pulse's $J_x$ in Figure 2b.[43,44] Accordingly, our Pt/FM1/Ta/FM2 synapses have potential novel advantages with versatile electrically controllable plasticity forms, such as variable STDP windows, the form of which can be modulated not only by altering the spike waveforms, but also potentially by electrically controlling the interlayer exchange couplings. This diversity in controllable STDP windows would serve for specific computing at different stages of the neuromorphic pathway (see the simulations in Supplementary S6 for detail).[42,45]

In conclusion, we have tuned a binary perpendicular ferromagnet FM1 into a multi-state spintronic Pt/FM1/Ta/FM2 artificial synapse that functions primarily on the basis of field-free spin-orbit torque driven domain wall motion in the FM1 layer with inherent in-plane interlayer exchange coupling from the FM2 layer. The multi-magnetic-state feature of FM1 is confirmed to be dependent on the in-plane coupling field. The ability to set multiple magnetic states, depending on the pulse number and magnitude enables spintronic implementations of plastic functionalities including EPSP, IPSP, and STDP. Our results suggest a practical way towards spintronic synaptic emulation using current controlled multi-state perpendicular magnetic materials with built-in equivalent in-plane magnetic fields, as artificial synapses for neuromorphic computing.

**Acknowledgements**

This work was supported by National Key R&D Program of China No.2017YFA0303400 and 2017YFB0405700. This work was supported also by the NSFC Grant No. 11474272, and 61774144. The Project was sponsored by Chinese Academy of Sciences, grant No. QYZDY-SSW-JSC020, XDB28000000 and XDPB12 as well.

**Author contributions**

K.W. conceived and designed the experiments. Y.C. fabricated and measured the devices. Y.S. assisted in measurement. A.R. and K.W. performed the theoretical analysis and modeling. A.R. and H.Z. discussed the results and commented on the manuscript. Y.C. and K.W. wrote the manuscript. All authors revised the manuscript.

**Additional information**

Correspondence and requests for materials should be addressed to K.W.

**Competing interests**

The authors declare no competing financial interests.

## Methods

**Film Preparation.** The films were deposited at room temperature onto Si wafers with a natural oxidation layer. Radio-frequency (RF) magnetron sputtering was used to deposit the AlO$_x$ layer, and d.c. magnetron sputtering was used to deposit the other layers. The base pressure of the chamber was less than $2\times10^{-6}$ Pa, and Ar gas was used for sputtering. The pressure of the chamber is $1.067\times10^{-1}$ Pa during deposition. No magnetic field was applied during the sputtering. The deposition rates for Ta, Pt, Co, AlO$_x$ layers were controlled to be ≈0.0139 nm s$^{-1}$, 0.0194 nm s$^{-1}$, 0.0093 nm s$^{-1}$, and 0.0016 nm s$^{-1}$, respectively.

**Device fabrication.** The deposited films were processed into Hall devices by two steps of standard electron-beam lithography (EBL) and Ar ion etching. First, the film was patterned into a cross shape Hall bar. Then, the central square area of the cross was patterned into a pillar, leaving the rest of the Hall bar with only stacks of Ta(0.5 nm)/Pt(3 nm).

**Characterization and measurement.** The Kerr characterization of the magnetization hysteresis and domain patterns was taken using a NanoMoke3 magneto-optical Kerr microscopy. The anomalous Hall effect measurements were carried out at room temperature with Keithley 2602B as the current source and Keithley 2182 as the nanovoltmeter. The input spikes for STDP measurement as shown in Figure 2c and 2e were generated using an Agilent B1500A semiconductor device analyzer with the semiconductor pulse generator unit (SPGU).

**Data availability.** The data that support the results of this study are available from the corresponding author on reasonable request.

**Supplementary Information:**

# Tuning a binary ferromagnet into a multi-state synapse with spin-orbit torque induced plasticity

Yi Cao[1], Andrew Rushforth[2], Yu Sheng[1], Houzhi Zheng[1,3], Kaiyou Wang[1,3,*]

[1]*State Key Laboratory for Superlattices and Microstructures, Institute of Semiconductors, Chinese Academy of Sciences, Beijing 100083, P. R. China*

[2]*School of Physics and Astronomy, University of Nottingham, Nottingham NG7 2RD, United Kingdom*

[3]*College of Materials Science and Opto-Electronic Technology, University of Chinese Academy of Science, Beijing 100049, P. R. China*

\* Corresponding emails: kywang@semi.ac.cn

**Table of Contents:**

## S1. Magnetic properties of the Pt/FM1/Ta/FM2 sample

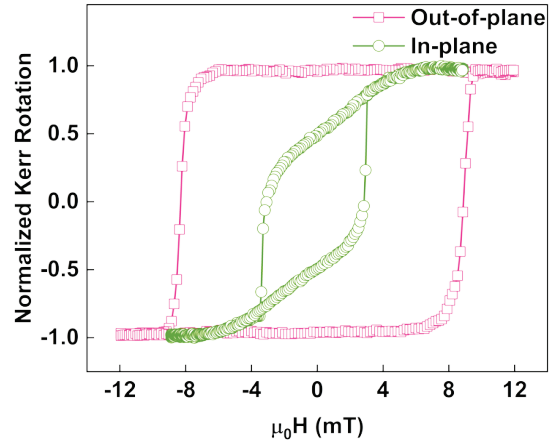

**Supplementary Figure 1 | Magneto-optic Kerr loops for the pillar of the Pt/FM1/Ta/FM2 double magnetic layered device.** The pink squares and the green circles correspond to Kerr loops measured out-of-plane and in-plane, respectively.

## S2. Adjustable current-induced magnetic switching direction by reversing the pre-magnetization of FM2

After applying a field of 20 mT along the -$x$-direction, the magnetization of FM2 was expected to be reversed. Then we removed the external field and measured the pulse current induced $R_H$-$J_x$ loop using the same pulse sequence that used for Manuscript Figure 1c. The result is shown in Supplementary Figure 2, where a clockwise $R_H$-$J_x$ loop indicates an $x$-directed interlayer exchange coupling field generated by the FM2 layer that magnetized along -$x$-direction. Note that the EPSP and IPSP here are induced by the negative and the positive increasing $J_x$, respectively, which are opposite to the EPSP and IPSP in Manuscript Figure 1c. In this way, the direction of synaptic plasticity in our device can be simply adjusted by controlling the magnetization of the FM2 layer.

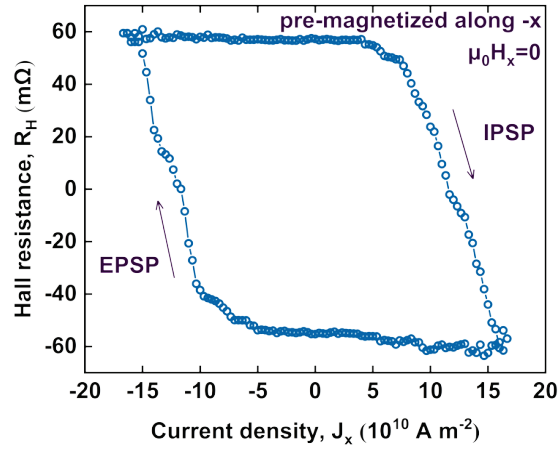

**Supplementary Figure 2 | Magnetic-field-free $R_H$-$J_x$ loop for the Pt/FM1/Ta/FM2 double magnetic layered device where the magnetization of FM2 layer is along -$x$-direction.** An -$x$ directed external magnetic field of 20 mT is applied to pre-magnetize the FM2 layer. Then the in-plane external field is removed. A sequence of pulses with scanning magnitude from -16.67×10$^{10}$ A m$^{-2}$ to 16.67×10$^{10}$ A m$^{-2}$ and then back to -16.67×10$^{10}$ A m$^{-2}$ is applied. The duration of every pulse is 10 ms, and every $R_H$ is measured 2 seconds after the application of each pulse. For the AHE detection of $R_H$, a small +$x$ directed d.c. current with fixed magnitude of 100 μA is used as the current source.

## S3. Magnitude of the interlayer exchange coupling field ($\mu_0 H_x^c$)

It is widely known that the magnitude and sign of interlayer exchange coupling oscillates with the thickness of the spacer layer between two magnetic layers.[46] In our Pt/FM1/Ta/FM2 system with the optimized stack structure of Ta(0.5)/Pt(3)/Co(1.3)/Ta(1.2)/Co(4)/AlO$_x$(2) (thickness in nm), the interlayer exchange coupling field ($\mu_0 H_x^c$) acting on the Co(1.3 nm) free layer coming from the upper Co(4 nm) fixed layer via the Ta spacer can be obtained from Supplementary Figure 3, where the $R_H$-$J_x$ loops are reversed at $\mu_0 H_x \approx$ 180 mT. Particularly, as $\mu_0 H_x$ increases from low value to 180 mT, the $R_H$-$J_x$ loops are anticlockwise; When $\mu_0 H_x >$ 180 mT, the R$_H$-J$_x$ loops become clockwise because the $\mu_0 H_x^c$ is overcome by the $\mu_0 H_x$, which results in a net magnetic field along +$x$-direction. Therefore, we have an antiferromagnetic interlayer exchange coupling with $\mu_0 H_x^c \approx$ -180 mT.

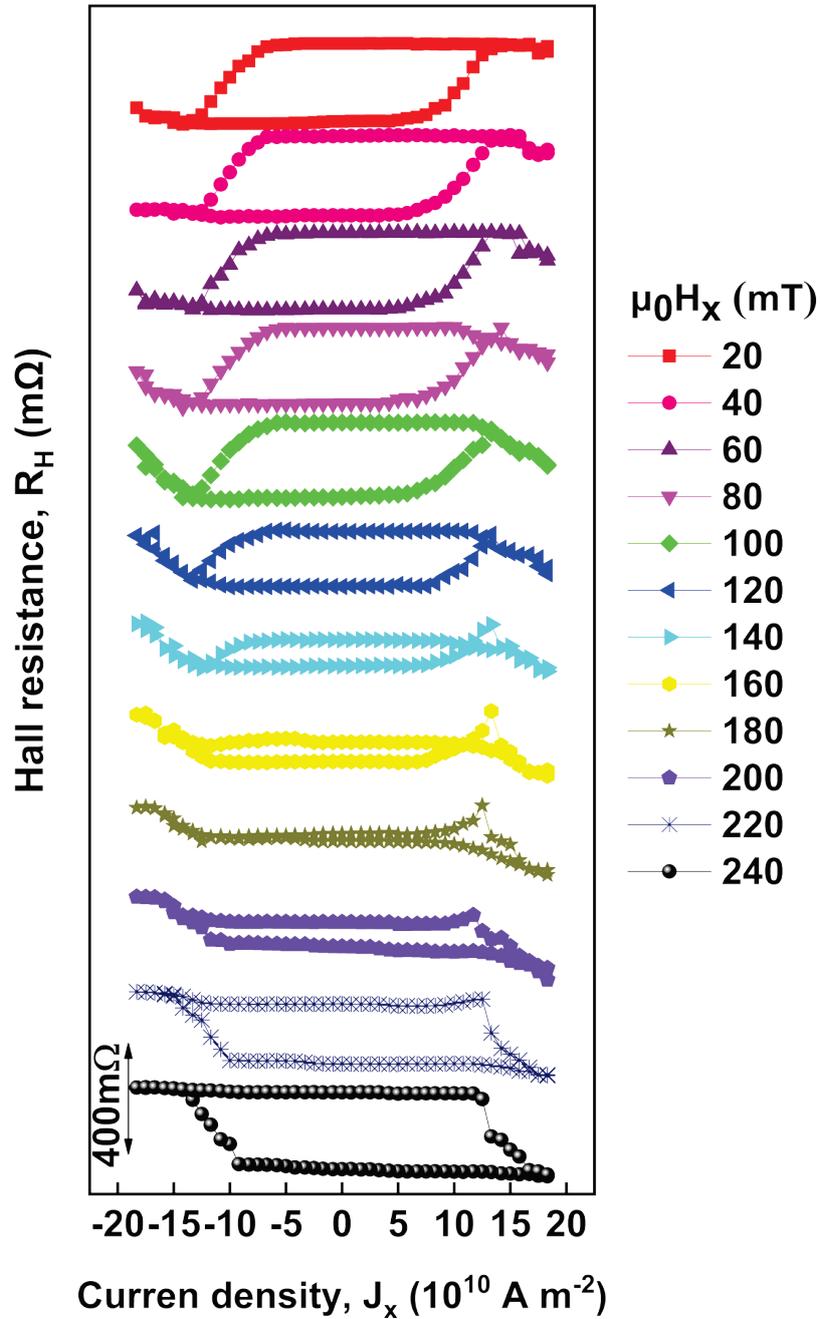

**Supplementary Figure 3 | Pulse current-driven $R_H$-$J_x$ SOT switching loops for the Pt/FM1/Ta/FM2 double magnetic layered device.** Loops with various $\mu_0 H_x$ from 20 mT to 240 mT are shifted vertically for clarity. The magnetization of FM2 Co(4 nm) layer is pre-magnetized along the +$x$-direction. The applied pulse sequence is a pulse string with each duration of 10 ms and scanning magnitude between the maximum ±$J_x$, and all $R_H$ data point are measured 2 seconds after the application of each pulse. For the AHE detection of $R_H$, a small +$x$ directed DC current with fixed magnitude of 100 μA is used as the current source.

**S4. Simulations of STDP resulted by overlapped spikes**

To simulate the STDP window by the overlapped spikes, we firstly fit the function $R_H^{single\ pulse}(J_x)$ of $R_H$ versus single pulse's $J_x$ from the experimental data in Manuscript Figure 2b. Then the amplitude of pre- and post-spikes ($J_{pre}$ and $J_{post}$) can be presented as time *t*-dependent piecewise functions. Because the post-spike shares the same waveform with the pre-spike but comes from the contrary direction with a time delay of $\Delta t$, thus we can overlap the pre- and post-spikes and obtain the amplitude of overlapped spikes ($J_{ovlp}$) as a function of $\Delta t$ by $J_{ovlp}(\Delta t) = J_{pre}(t) - J_{post}(t, \Delta t)$. Finally, the STDP window (in form of a $R_H$-$\Delta t$ function) can be obtained by plugging the $J_{ovlp}$-$\Delta t$ function into the $R_H^{single\ pulse}$-$J_x$ function, which can be represented as $R_H^{STDP}(\Delta t) = R_H^{single\ pulse}\left(J_{ovlp}(\Delta t)\right)$.

Using the experimental waveform function of E0, E1 and E2 in Manuscript Figure 2c and 2e, the simulated STDP windows, as shown in Supplementary Figure 4b, are very similar to that of the experimental results. If we change the decay part of the spike waveform to an exponential function, as shown in Supplementary Figure 4c, the exponential STDP window postulated by Bi and Poo can also be obtained, as shown in Supplementary Figure 4d.[42] Thus, by modifying the spike waveform, different STDP windows can be obtained, which could enable our synaptic device with the capabilities of different learning rules.[45]

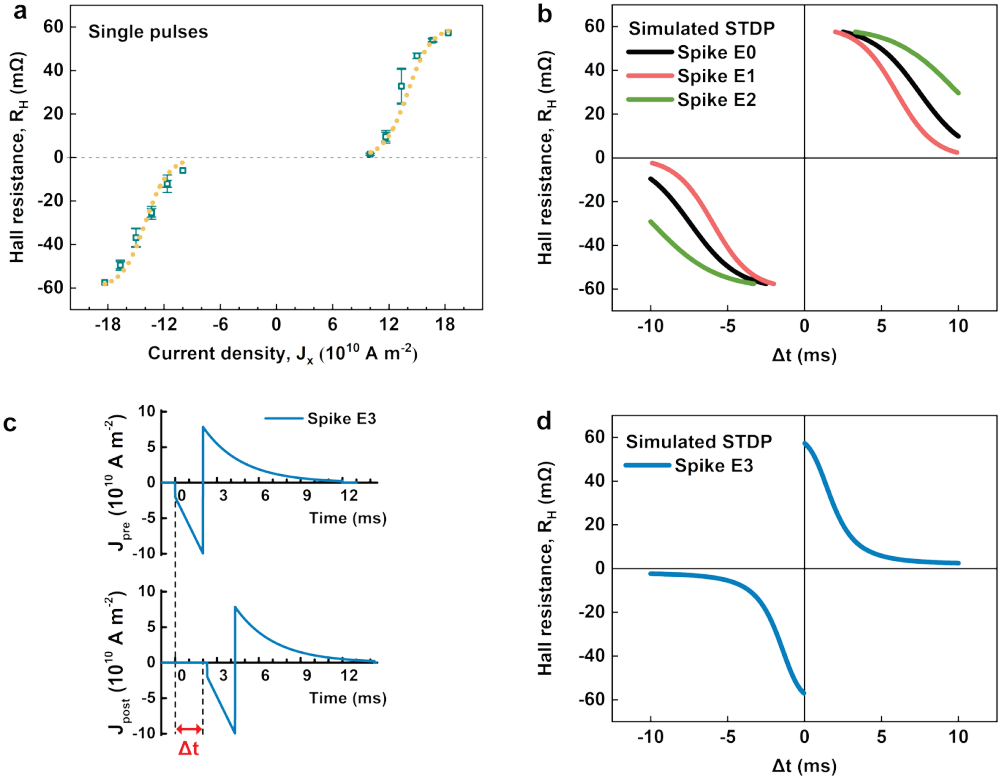

**Supplementary Figure 4 | Simulations of STDP resulting from overlapped spikes. a,** Fittings (yellow dotted line) of the $R_H$ changes by single pulses in Manuscript Figure 2b. **b,** simulated STDP windows for Spike E0, E1, and E2. **c,** Spike waveform with exponentially decaying amplitude (Spike Exp). **d,** simulated STDP window for Spike Exp.

## S5. The magnetization reversal mechanism

### 5.1 Domain wall motion images

We directly imaged the domain patterns of the Pt/FM1/Ta device with 20 μm-wide squared pillar by magneto-optic Kerr mapping. During the measurements, an external in-plane field of $\mu_0 H_x$ = 120 mT was applied, where the parallelogram-like current-induced magnetic switching loop in the Pt/FM1/Ta reference sample is similar to that of the Pt/FM1/Ta/FM2 system. Supplementary Figure 5a shows the current-driven $R_H$-$J_x$ loop, where a sequence of pulses with scanning magnitude from $4.5\times10^{10}$ A m$^{-2}$ to -$4.5\times10^{10}$ A m$^{-2}$ and then back to $4.5\times10^{10}$ A m$^{-2}$ was applied. The duration of each pulse was 10 ms and the interval between neighboring pulses was 2 seconds, within which time more than one Kerr image could be taken. The corresponding Kerr microscope images of the pillar during its down-to-up and up-to-down $m_z$ switching processes are shown in Supplementary Figures 5b1-b8 and Supplementary Figures 5c1-c8, respectively. At the beginning of the down-to-up switching process with threshold pulse magnitude $J_x = -2\times10^{10}$ A m$^{-2}$, nucleation of a small domain is found to appear at the edge of the pillar, as shown by the dark area (up $m_z$) in Supplementary Figure 5b1. The domain wall propagates across the pillar with the increasing negative $J_x$ (Supplementary Figures 5b2-b7), and finally fills up the pillar area at a finishing switching current $J_x = -3.75\times10^{10}$ A m$^{-2}$ (Supplementary Figure 5b8). It is similar for the up-to-down switching process, where nucleation of two or three domains form at the current-entrance edge (bright areas in Supplementary Figure 5c1) before they enlarge and merge as they propagate across the pillar along the $+x$-direction as $J_x$ increases (Supplementary Figures 5c2-c9). Regardless of the nucleation numbers, the switching process in our Pt/FM1/Ta structure with $\mu_0 H_x = 120$ mT is thus dominated by domain wall propagation. Thus, the underlying mechanism for the multi-state magnetization reversal in our Pt/FM1/Ta structure with in-plane external field, and therefore the Pt/FM1/Ta/FM2 structure with inherent antiferromagnetic coupling field, should be ascribed to a pulse magnitude-dependent domain wall motion process.

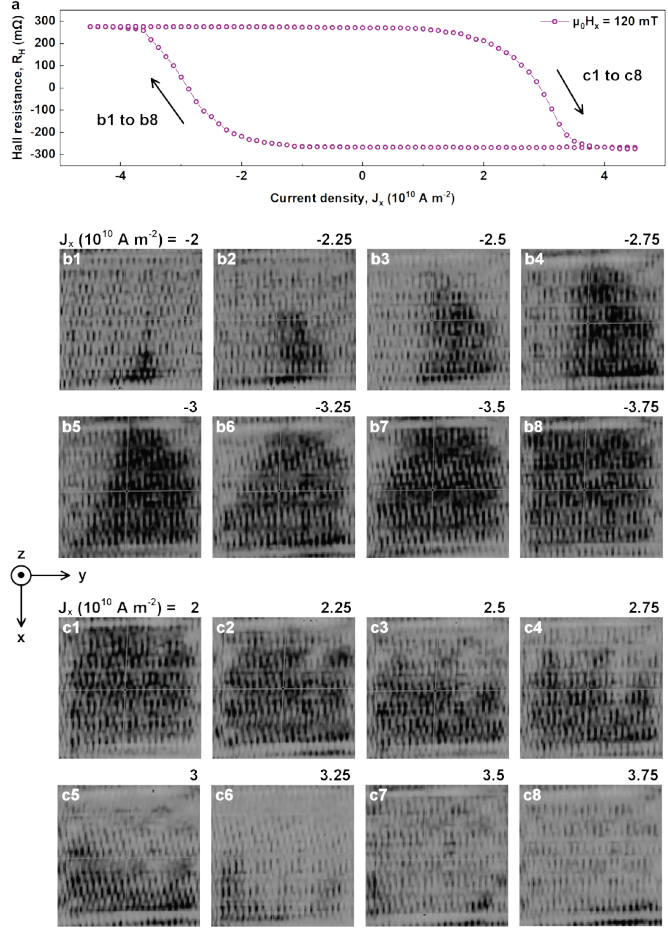

**Supplementary Figure 5 | Current-driven domain wall motion in a 20 μm-wide Pt/FM1/Ta squared pillar with $\mu_0 H_x$ = 120 mT. a,** Pulse current-driven $R_H$-$J_x$ loop with external magnetic field $\mu_0 H_x$ = 120 mT. The sequence of pulses each with duration of 10 ms and scanning magnitude from $4.5 \times 10^{10}$ A m$^{-2}$ to $-4.5 \times 10^{10}$ A m$^{-2}$ and then back to $4.5 \times 10^{10}$ A m$^{-2}$ is applied, and all $R_H$ data points are measured 2 seconds after the application of each pulse. **b1-b8,** Kerr microscope images during the down-to-up switching process from negative $R_H$ to positive $R_H$. **c1-c8,** Kerr microscope images during the up-to-down switching process from positive $R_H$ to negative $R_H$. Bright and dark contrast corresponds to down and up $m_z$ directions, respectively.

## 5.2 Determination of the DMI field

Current induced domain wall motion in ferromagnetic metal (FM)/ heavy metal (HM) systems with perpendicular magnetic anisotropy is understood to arise from a combination of the spin transfer torque (STT) mechanism due to electrons flowing within the ferromagnetic material, and the anti-damping (also known as Slonczewski-like) torque due to the spin accumulation at the heavy metal/ferromagnet interface, generated by the spin Hall effect (SHE) in the heavy metal[47]. We refer to the later mechanism as SHE-STT to distinguish it from the former conventional STT mechanism. The Rashba effect, arising due to an asymmetric layer structure, can give rise to field-like[48] and Slonczewski-like torques[40,49]. Finally, the Dzyaloshinskii-Moriya interaction (DMI) arising at the heavy metal/ferromagnet interface stabilises the Néel domain wall configuration with the same preferred chirality for up/down and down/up domain wall configurations[50]. Typically, the SHE-STT provides the dominant torque, leading to the motion of up/down and down/up domain walls in the same direction when the DMI ensures that both domain walls possess the same chirality. This property prevents the magnetization reversal of devices by SHE-STT unless the DMI is compensated by an external effective magnetic field so that up/down and down/up domain walls possess opposite chirality and can be driven in opposite directions by the current.

The magnitude of the field arising due to the DMI can be determined by measuring the d.c. current $J_x$ induced perpendicular effective field ($\mu_0 H_z^{eff}$) as a function of the applied in-plane bias magnetic field ($\mu_0 H_x$) until $\mu_0 H_x$ approaches the DMI field. $\mu_0 H_z^{eff}$ can be obtained by measuring the magnetic hysteresis loops under a particular $J_x$ and $\mu_0 H_x$, where the loop for a positive (negative) $\mu_0 H_x$ will offset leftward (rightward) with a value of $\mu_0 H_z^{eff}$ from the loop for $\mu_0 H_x = 0$.[51] Following this method, we measured the out-of-plane MOKE loops at the pillar of our 6 μm-wide Pt/FM1/Ta device with a fixed d.c. $J_x = 1.39 \times 10^{10}$ A m$^{-2}$ and various $\mu_0 H_x$, as shown in Supplementary Figure 6a. The obtained $\mu_0 H_z^{eff}$ as a function of the $\mu_0 H_x$ is shown in Supplementary Figure 6b. From this we can estimate that the strength of the field due to the DMI in our devices is approximately 40 mT.

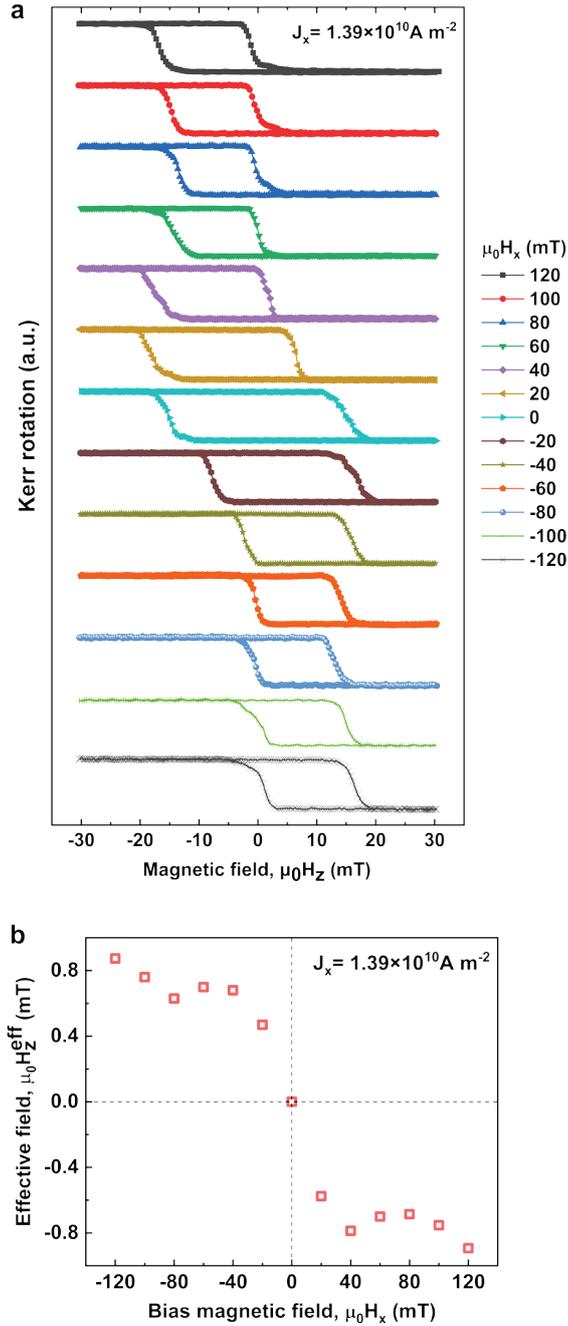

**Supplementary Figure 6 | D.C. current $J_x$ induced perpendicular effective field $\mu_0H_z^{eff}$ for the Pt/FM1/Ta reference device under various in-plane bias magnetic field $\mu_0H_x$.** The pillar width of the device is 6 μm. **a**, Out-of-plane MOKE loops measured under a fixed d.c. current with $J_x = 1.39\times10^{10}$ A m$^{-2}$ and various in-plane bias magnetic field $\mu_0H_x$. Loops are shifted vertically for clarity. The relative horizontal offset between each loop and the loop for $\mu_0H_x = 0$ can be read as the current induced perpendicular effective field $\mu_0H_z^{eff}$, whose evolution with the $\mu_0H_x$ is plotted in **b**.

## 5.3 Numerical calculations

We modelled the motion of a domain wall using the rigid one-dimensional model modified to include the effects of STT, SHE-STT and DMI. The model has been shown to provide a good qualitative description of domain wall motion along nanowire strips with perpendicular magnetic anisotropy, including the effects of the SHE-STT and DMI[52-55]. The model describes the domain wall behaviour through two coupled Equations (1) and (2) based on the position $X$ along the nanowire and the azimuthal angle $\phi$.

$$(1 + \alpha^2)\frac{dX}{dt} = \alpha\gamma\Delta H_Z + (1 + \alpha\beta)b_J + \frac{\alpha\gamma\Delta\pi}{2}H_{SHE}\cos(\phi) +$$
$$\frac{\gamma\Delta\pi}{2}H_{DMI}\sin(\phi) + \frac{\gamma\Delta\pi}{2}H_x\sin(\phi) \quad (1)$$

$$(1 + \alpha^2)\frac{d\phi}{dt} = \gamma H_Z + (\beta - \alpha)\frac{b_J}{\Delta} + \frac{\gamma\pi}{2}H_{SHE}\cos(\phi) - \frac{\alpha\gamma\pi}{2}H_{DMI}\sin(\phi) -$$
$$\frac{\alpha\gamma\pi}{2}H_x\sin(\phi) \quad (2)$$

Where the parameters, with values appropriate for the studied Pt/Co/Ta trilayer system include the Gilbert damping term, $\alpha = 0.2$, the gyromagnetic ratio, $\gamma = 2.21 \times 10^5$ m A$^{-1}$ s$^{-1}$, the domain wall width, $\Delta = 10$ nm, and the STT non-adiabaticity parameter, $\beta = 0$. STT is represented by $b_J = \mu_B P j / e M_S$, where $j$ is the current density, $P = 0.5$ is the spin polarisation, $M_S = 1.4 \times 10^6$ A m$^{-1}$ is the magnetization of the Co layer, $\mu_B$ is the Bohr magneton and $e = -1.6 \times 10^{-19}$ C is the charge on the electron. The effective fields due to the SHE and DMI are $H_{SHE} = \hbar\theta_{SH}j/2\mu_0 e M_S t$, and $H_{DMI}$, in the directions transverse ($y$) and parallel ($x$) to the current direction respectively. $\theta_{SH} = 0.1$ is the spin Hall angle and $t = 1.3$ nm is the Co layer thickness. $H_x$ is the effective magnetic field in the plane of the layer in the $x$-direction. We set $H_{DMI} = 0$ for the calculations because it is clear that $H_{DMI}$ only acts as an offset to $H_x$.

We modelled the effects of pinning as a spatially dependent field component in the z-direction given by $H_Z = B\frac{\partial V}{\partial X}$, where $B$ represents the strength of the pinning potential and $V = sin^2(\pi X/\xi)$ with $\xi = 30$ nm. Thermal effects were modelled as an uncorrelated Gaussian distributed stochastic fluctuation of $H_Z$. Each calculation was averaged over five such random distributions.

Supplementary Figure 7a shows the calculated domain wall velocity, $v$, versus the current density $j$, for various $\mu_0 H_x$, when the pinning potential is relatively weak ($B = 8\times10^{-29}$ J). As current increases, the velocity increases initially, before saturating at some value which is dependent on $\mu_0 H_x$. This is because the SHE-STT term tends to drive the azimuthal angle, $\phi$, towards some equilibrium value between the Bloch and Néel configurations, at which point $d\phi/dt = 0$ and a constant velocity is maintained which depends on the balance between the terms involving $H_{SHE}$ and $H_x$. For larger $\mu_0 H_x$, the velocity is higher at any value of current because the applied field acts against the tendency for $\phi$ to rotate towards the Bloch configuration, thereby allowing the SHE-STT to make a more significant contribution to the overall velocity. In this regime, the pinning potential has a relatively weak effect on the domain wall motion.

Supplementary Figure 7b shows the calculated velocity versus current for a relatively strong pinning potential ($B = 8\times10^{-28}$ J). Now the pinning potential has a more significant effect on the domain wall motion. For low current densities the velocity is suppressed because the pinning potential prevents uniform domain wall motion. At higher current densities the domain wall motion becomes less stochastic and the velocity increases significantly with increasing current density, until it saturates in a similar manner to that observed for the weak pinning potential.

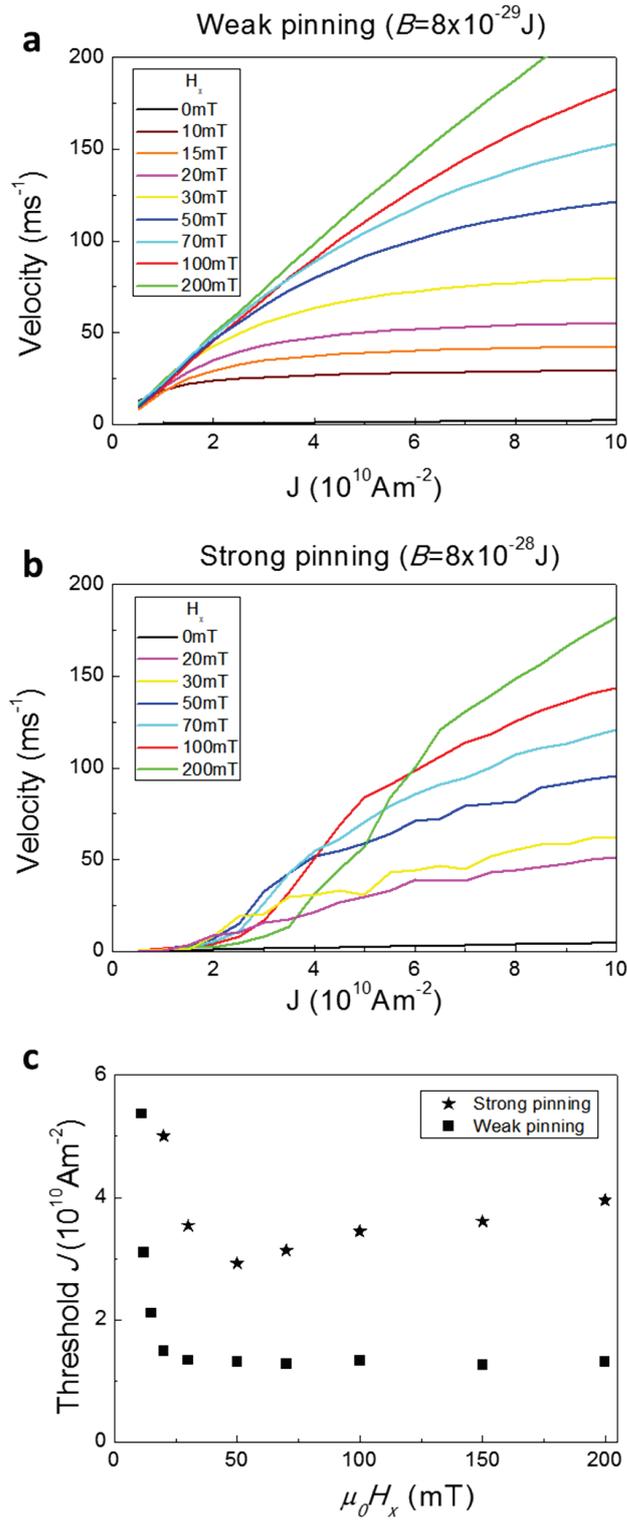

**Supplementary Figure 7 | Calculated domain wall velocity, plotted against current density, $j$, for various in-plane magnetic field $\mu_0H_x$.** Calculated velocity for **a,** weak and **b,** strong pinning potentials. **c,** the calculated threshold current density versus $\mu_0H_x$ for the weak and strong pinning regimes.

## 5.4 Discussions of the broadened magnetization reversal process

At low current densities, where domain wall motion is stochastic, increasing $\mu_0 H_x$ decreases the importance of the terms involving $H_{SHE}$ relative to those involving $H_x$ in Equations (1) and (2), and $\phi$ remains close to the Néel configuration i.e. the SHE-STT term does not drive $\phi$ significantly away from $\phi = 0$. Domain wall motion can only proceed by stochastic fluctuations of $\phi$, that are large enough to allow the last term in Equation (1) to drive the domain wall motion across the pinning potential landscape. Increasing $\mu_0 H_x$ supresses the fluctuations in $\phi$ and therefore reduces the average domain wall velocity in this regime.

At high current densities, the opposite trend is observed. Here, $H_{SHE}$ is significantly larger than $H_x$ and is the dominant term in determining the domain wall velocity. As in the case with weak pinning potential, increasing $\mu_0 H_x$ causes the equilibrium value of $\phi$ to be closer to the Néel configuration, thereby increasing the contribution of the SHE-STT term to the domain wall velocity. Consequently, at high current densities, increasing $\mu_0 H_x$ causes the domain wall velocity to increase.

At intermediate current densities there is a transition between the regimes described above. For low $\mu_0 H_x$, increasing $\mu_0 H_x$ causes domain wall velocity to increase, whereas for higher $\mu_0 H_x$, the opposite trend is observed. This intermediate regime gives a good qualitative agreement with our experimental observations. We define a threshold domain wall velocity that is necessary in order to observe switching of the magnetization (in this case we choose $v = 30$ m s$^{-1}$, but the absolute value is not too important because the model is expected to give only qualitative understanding) and we plot the current density at which that velocity is reached as a function of $\mu_0 H_x$ in the weak and strong pinning regimes. As shown in Supplementary Figure 7c, the threshold current density for the weak pinning regime decreases with increasing $\mu_0 H_x$ in a manner similar to the experimental $J_x^{th}$, while the threshold current density for the strong pinning regime initially decreases with increasing $\mu_0 H_x$ before increasing for further increases in $\mu_0 H_x$ in a manner similar to the experimental $J_x^f$. This suggests that there is a distribution of strong and weak pinning sites in our device and that the onset of the magnetization reversal at $J_x^{th}$ occurs when the current drives the domain wall across weaker pinning sites, while the magnetization reversal completes at values of $J_x^f$ sufficiently large to drive the domain wall across the stronger pinning sites. The different efficiencies of the spin-orbit torque in driving domain wall motion in the strong and weak pinning regimes leads to the broadening of the magnetic reversal transition as $\mu_0 H_x$ is increased.

The magnetic field applied in the plane will also tilt the magnetization in the regions of up and down pointing domains with the effect of reducing the magnitude of the z-component of the magnetization. This effect is not captured within the 1-dimensional domain wall model, so the possible effect on the domain wall velocity is not clear. It may account for the decrease of $J_x^f$ at the highest values of $\mu_0 H_x$. It is likely to be responsible for the observed reduction of the magnitude of the change of $R_H$ when the magnetization switches at high values of $\mu_0 H_x$.

Finally, we note that for current to flow within the FM Co layer in the pillars, it must first enter the region via the Pt layer. Therefore, the current density in the Pt will be largest at the edges where it enters or leaves the region of the pillar. The SHE-STT will be largest in these regions, which may explain why the domain walls observed in Supplementary Figure 5 nucleate at one edge of the pillar and propagate towards the opposite end.

**S6. Simulations of STDP under different magnitude of interlayer exchange coupling**

According to Manuscript Figure 3, all the $J_x^{th}$, the $J_x^f$, and the $\Delta J_x$ of the Pt/FM1/Ta sample change remarkably with the in-plane external field. Hence it is obvious that the magnitude of the interlayer exchange coupling field in the Pt/FM1/Ta/FM2 device would affect the switching parameters significantly, including the function of $R_H$ versus single pulse's $J_x$ in Manuscript Figure 2b.

Here we set two different functions of $R_H$ versus single pulse's $J_x$, as shown in Supplementary Figure 8a, for two different supposed coupling magnitudes (C1 and C2). Then we use the same exponentially decaying spike waveform function E3 in Supplementary Figure 4c to simulate the $R_H$-$\Delta t$ functions for C1 and C2, respectively, and obtained two different STDP windows with distinctive decay slopes as a function of the |Δ*t*|, as shown in Supplementary Figure 8b. This simulation shows that the plasticity of our Pt/FM1/Ta/FM2 can be efficiently controlled by means of modulating the magnitude of its interlayer exchange coupling field, which would bring more versatile computing capabilities in addition to the spike-waveform-dependent STDP window forms.

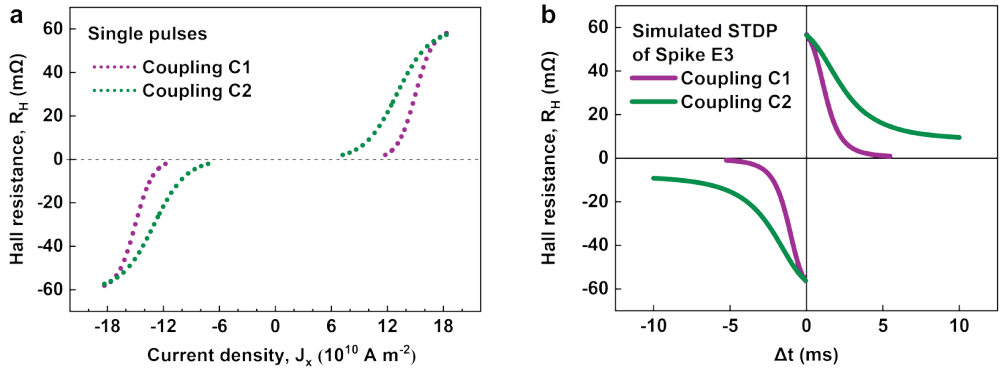

**Supplementary Figure 8 | Simulations of STDP under different functions of $R_H$ versus single pulse's $J_x$. a,** The different given functions of $R_H$ versus single pulse's $J_x$ (for two corresponding supposed coupling magnitudes, C1 and C2, respectively). **b,** Simulated STDP windows for C1 and C2 with spike waveform in Supplementary Figure 4d (Spike Exp).